\documentclass[pdftex,12pt,a4paper]{article}
\usepackage{jheppub} 

\usepackage{multicol}

\usepackage{enumerate}
\usepackage{amsmath,amsfonts,amssymb,amscd,amsthm}
\usepackage{amsfonts}

\usepackage[english]{babel}
\usepackage[latin1]{inputenc}

\usepackage{slashed}
\usepackage{graphicx}

\usepackage{stackrel}
\usepackage{tabularx}
\usepackage{color}
\usepackage[T1]{fontenc}

\usepackage{lineno}

\newcommand{\roughly}[1]{\mathrel{\raise.3ex\hbox{$#1$\kern-0.85em
\lower1ex\hbox{$\sim$}}}}

\def\be{\begin{equation}}
\def\beq\begin{equation}
\def\bea{\begin{eqnarray}}
\def\eea{\end{eqnarray}}

\def\beq{\begin{equation}}
\def\eeq{\end{equation}}
\def\beqa{\begin{eqnarray}}
\def\eeqa{\end{eqnarray}}

\def\dsl{\hbox{/\kern-.5300em$\partial$}}

\newcommand{\bmat}{\left(\begin{array}}
\newcommand{\emat}{\end{array}\right)}

\def\-{\hphantom{-}}

\def\s2{\frac{1}{2}}

\def\IF{\relax{\rm I\kern-.18em F}}
\def\II{\relax{\rm I\kern-.18em I}}
\def\IP{\relax{\rm I\kern-.18em P}}
\def\IC{\relax{\rm I\kern-.48em C}}
\def\IK{\relax{\rm I\kern-.20em K}}
\def\IM{\relax{\rm I\kern-.25em M}}

\def\Dsl{\,\raise.15ex\hbox{/}\mkern-13.5mu D} 
\def \one{\relax{\rm 1\kern-.26em I}}

\newcommand\g{\gamma}

\newcommand\kl{K_L}

\newcommand\klm{K^M_L}

\newcommand\kslinv{K_S,K_L \to invisible}

\newcommand\ksl{K_{S,L} \to invisible}
\newcommand\ksinv{K_S \to invisible}
\newcommand\klinv{K_L \to invisible}
\newcommand\koinv{K^0 \to invisible}

\newcommand\kob{\overline{K}^0}

\definecolor{darkerblue}{rgb}{0.0,0.0,0.5}
\newcommand{\seq}{\begin{subequations}}
\newcommand{\sen}{\end{subequations}}

\newcommand{\eq}{\begin{eqnarray}}
\newcommand{\en}{\end{eqnarray}}

\begin{document}

\title{Search for  $K_{S,L}$ oscillations and invisible decays  into the dark sector at NA64}

\date{\today}

\author{
S.N.~Gninenko$^{1,2}$, N.V.~Krasnikov$^{1,2,3}$, V.A.~Matveev$^{1,2}$}

\affiliation{$^{1}$ Institute for Nuclear Research of the Russian Academy of Sciences, 117312 Moscow, Russia \\
  $^{2}$ Joint Institute for Nuclear Research, 141980 Dubna, Russia,\\
  $^3$ Corresponding author}

\abstract{
The  decays $\ksl$ have never been experimentally tested. In the Standard Model their  branching ratios for the decay into two neutrinos are predicted to be extremely small, $Br(K_{S,L} \to \nu \bar{\nu}) \lesssim 10^{-16}$. We consider several natural extensions  of the SM, such as two-Higgs-doublet (2HDM), 2HDM and light scalar, and dark mirror  sector models, that allow to enhance the  $Br(\ksl)$  up to a measurable level. We briefly discuss the possible  search for $\ksl$ decays and $K_{S,L}$  oscillations into the dark sector at  the  
NA64 experiment at CERN with the sensitivity to $Br(\ksl) \lesssim 10^{-7}-10^{-5}$.

$$ $$

keywords: physics beyond standard model, kaon physics, fixed target experiments

}




\maketitle
\addcontentsline{toc}{section}{Table of Contents}

\newpage

\section{Introduction}
\label{sec:intro}
Experimental studies of   invisible decays, i.e. particle transitions to 
an experimentally unobservable final state, played an important role both in the development of the
standard model (SM) and in testing its extensions \cite{pdg}. 
It is worth remembering the precision measurements of the  $Z \to invisible$ decay rate at LEP  for the determination of the number of lepton families in the SM.  
In recent years, experiments on invisible particle decays have  received considerable attention.
Motivated by  various models of physics beyond the SM, see, e.g. Refs.~\cite{tulin, kam,Gninenko,GK2015,GK2016,gabri,host,redtop,toro,aszh,aszh1,aszh2,vor} and references therein,  these experiments include searches for 
invisible decays of $\pi^0$ mesons at  E949  \cite{pi0} and NA62 \cite{na62}, $\eta$ and $\eta'$ mesons
 at BES \cite{bes14} and NA64\cite{na642024},
  heavy $B$-meson decays at Belle \cite{belle}, BaBAR \cite{babar}, and  BES \cite{bes},  and
 invisible decays of the upsilon(1S) resonance at CLEO \cite{cleo},  baryonic number violation with  nucleon disappearance at SNO \cite{sno}, BOREXINO \cite{borexino}, and  KamLAND \cite{kamland}, see also Ref.\cite{tretyak}, electric charge-nonconserving electron decays $e^- \to invisible$  \cite{klap},   neutron-mirror-neutron oscillations at PSI \cite{psinn1,psinn2} and  the ILL reactor \cite{ser}, and the disappearance of neutrons into another brane world \cite{sar}. 
One could also mention experiments looking for extra dimensions and  dark mirror matter through the 
invisible decays of positronium \cite{gkr,bader, paolo}, and plans for new   experiments to  search for muonium annihilation into two neutrinos,
 $\mu^+ e^- \to \nu \overline{\nu}$ \cite{muonium},  and electric charge nonconservation in the muon decay  $\mu^+ \to invisible$  \cite{sngmu}.

\par The aim of this paper is to discuss an experiment with the NA64 detector at CERN  modified for the sensitive search for $\ksl$ decays.
 The rest of the paper is organized as follows. 
 In Sections 2, 3 and 4, we  briefly review the motivations to perform the search, and several natural extensions  of the SM, such as two-Higgs-doublet (2HDM), 2HDM and light scalar, and mirror dark sector models, that allow to  enhance the  $Br(\ksl)$  up to a measurable level. In Section 5  we discuss the search method and the
 NA64 setup modified for the searching for $\ksl$ decays, and  $K^0$ - dark $K^0$ oscillations. The background sources and the expected sensitivity  
are also discussed. Section 6 contains concluding remarks. 
 
\section{Motivations}
The decays $\ksl$ have never been experimentally tested despite the extensive search program of new physics in  kaon decays \cite{pdg}. The first  bound on $ Br(\ksl ) \lesssim 10^{-4}-10^{-3}$ 
has been set assuming validity of unitarity in the $K^0$ sector  from  the existing experimental data \cite{Gninenko}. From the experimental viewpoint, the $K$-mesons themselves have brought to the SM so many surprises that all their still unknown properties deserve to be carefully studied.
Since long ago it was recognized that  $\ksl$ decays "would be interesting to explore, but its detection looks essentially impossible. 
New ingenious experimental ideas are required"  \cite{marci}.   
\par One of approaches  proposed not long ago in Ref. \cite{Gninenko}, is based on the idea to use  charge-exchange reaction as a source of well-tagged neutral mesons.
In this process, the $\ksl$ events would exhibit themselves via a striking signature - the complete disappearance of the incoming beam energy in the detector. 
 The first results on the search for the $\eta, \eta' \to invisible$ decay modes recently obtained by the NA64 Collaboration at the CERN SPS \cite{na642024} provide
  proof-of-concept  and  suggest the overall future direction for performing such kinds of experiments  with  this approach.   

\subsection{General considerations}
The  $\ksl$ decays are complementary to the $K^{+} \to \pi^+ + invisible$ and  $K_L \to \pi^0 + invisible$ decays, whose 
 branching ratios in the Standard Model (SM)  are  predicted  to be 
\cite{buras}
\begin{equation}
 Br(K_L \rightarrow \pi^0 \nu \bar{\nu}) = (2.6 \pm 0.4) \times 10^{-11}  \,,
\label{klsm}
\end{equation}  
\begin{equation}
 Br(K^+ \rightarrow \pi^+ \nu \bar{\nu}) = (8.5 \pm 0.7) \cdot 10^{-11}\,, 
\label{k+sm}
\end{equation}
with the invisible final state represented by neutrino pairs. A powerful comparison between experiment and theory
is possible  due to the accuracy of both the measurements and the SM calculations of these observables. 
A discrepancy would signal the presence of physics beyond the Standard Model (BSM) making  
the precision study of these decays an effective probe to  search for it, see e.g. 
\cite{buras,ciri,bryman,kom,Monika,lgl}.

\par On the contrary, the searching for invisible  decays of the $K_S$ and  $K_L$, and other  pseudoscalar mesons ($M^0$), such as $\pi^0, \eta, \eta',$ is particularly  advantageous because in the SM the branching fraction  of their  decay into a  neutrino-antineutrino pair, ${\rm Br}(M^0\to  \nu \overline{\nu})$,  is predicted to be   extremely small~\cite{marci}. For massless neutrinos, this transition is forbidden kinematically by angular momentum conservation. Indeed, in the $M^0$ rest frame the neutrinos produced in the decay fly away in  opposite directions along the same line. Since the neutrinos and antineutrinos are
massless, the projection of the sum of their spins on this line equals $\pm$1. The projections of the orbital angular momentum of the neutrino on this line are equal to
zero. Since in the initial state we have a scalar, the process is forbidden. For the case of massive neutrinos, one of them is forced to  have the  "wrong" helicity resulting  in the suppression of  ${\rm Br}(M^0\to  \nu \overline{\nu})$ by  a factor proportional to  the neutrino mass squared,   
\begin{equation}
{\rm Br}(M^0\to  \nu \overline{\nu}) \sim \frac{m_\nu^2}{m_{M^0}^2} \lesssim 10^{-16} , 
\label{eq:invrate}
\end{equation}
for $m_\nu \lesssim 10$ eV and $m_{M^0} \simeq m_{K} \simeq  0.5$ GeV \cite{pdg}. In the SM the helicity suppression can be overcome for the four-neutrino final state, however,  in this case, ${\rm Br}(M^0 \to  \nu \overline{\nu} \nu \overline{\nu}) \lesssim 10^{-18}$~\cite{gao}. Therefore, differently form the decays \eqref{klsm},\eqref{k+sm}, observation of the $M^0\to invisible$  decay for any of $M^0$ mesons   would unambiguously signal the presence of BSM physics.

\subsection{The Bell-Steinberger unitary relation}
 Another important reason to look for $\kslinv$ decays is related to additional tests
of the $K^0-\overline{K}^0$ system using the Bell-Steinberger relation\cite{bs}. 
This relation, obtained by using the unitarity condition, connects
 CP and CPT violation in the mass matrix of the kaon system, i.e.  parameters describing T and CPT noninvariance,  to CP and CPT violation in all decay channels of neutral kaons,  see, e.g. Refs. \cite{js, adpdg, lm, dafne, bloch}. The CPT appears to be an exact symmetry of nature, while C, P and T are known to be violated. Hence, testing the validity of the CPT invariance  probes the basis of the SM. The Bell-Steinberger relation remains one of the most sensitive 
  tests of CPT symmetry, resulting,  for example,  to the impressive sensitivity of $-5.3\times  10^{-19}$ GeV $<m_{K^0} -  m_{\overline{K}^0}< 6.3\times 10^{-19}$ GeV at 95\% C.L. for the neural kaon mass difference \cite{kloe,cplear}.  However, the  question  of how much the invisible decays of $K_S$ or $K_L$ can influence the precision of the Bell-Steinberger analysis  still remains open \cite{worksh}. This makes the future searches for these decay modes  very  interesting and complementary to the study of other  $K_{S,L}$ decays.
\par Briefly,  within the Wigner-Weisskopf approximation, the time evolution of the neutral kaon system is described by \cite{kloe}:
\begin{equation}
i\frac{d\Phi(t)}{dt}=H\Phi(t)=\Bigl(M-\frac{i}{2}\Gamma\Bigr) \Phi(t)
\label{kk}
\end{equation}
where $M$ and $\Gamma$ are $2\times 2$  Hermitian matrices, which are time independent,  and $\Phi(t)$ is a two-component state vector in the $K^0 - \overline{K}^0$ space. Denoting by $m_{ij}$ and $\Gamma_{ij}$ the elements of $M$ and $\Gamma$ in the  $K^0 - \overline{K}^0$basis, $CPT$ invariance implies
\begin{eqnarray}
m_{11}=m_{22}~~ (\rm{or} ~m_{K^0}=m_{\overline{K}^0})~ \rm{and} \\ \nonumber
 \Gamma_{11}=\Gamma_{22}~~  (\rm{or} ~\Gamma_{K^0}=\Gamma_{\overline{K}^0})
\end{eqnarray}

The eigenstates of Eq. (\ref{kk}) can be written as
\begin{eqnarray}
K_{S,L}= \frac{1}{\sqrt{2(1+|\epsilon_{S,L}|^2)}}\Bigl((1+\epsilon_{S,L})K^0  \nonumber \\
\pm (1-\epsilon_{S,L})\overline{K}^0) \\ \nonumber
\end{eqnarray}
with
\begin{eqnarray}
\epsilon_{S,L}  = \frac{1}{m_L-m_S+i(\Gamma_S -\Gamma_L)/2}\Bigl[-i\rm{Im}(m_{12})-\\ \nonumber
\frac{1}{2}\rm{Im}(\Gamma_{12}) \pm \frac{1}{2}(m_{\overline{K}^0}-m_{K^0}-\frac{i}{2}(\Gamma_{\overline{K}^0}-\Gamma_{K^0})\Bigr] \equiv \epsilon \pm \delta 
\end{eqnarray}
The unitarity condition allows us  to express the four elements of $\Gamma$ in terms of appropriate combinations of the 
kaon decay amplitudes $A_i$:
\begin{equation}
\Gamma_{ij}=\sum_f A_i(f)A_j^*(f), ~~i,j=1,2=K^0,\kob
\end{equation}
where the sum is over all the accessible final states. 
\begin{eqnarray}
\Bigl(\frac{\Gamma_S+\Gamma_L}{\Gamma_S-\Gamma_L}+i\rm{tan}\phi_{SW}\Bigr) \Bigl(\frac{Re(\epsilon)}{1+|\epsilon|^2}-i\rm{Im}(\delta)\Bigr)  \nonumber \\
=\frac{1}{\Gamma_S-\Gamma_L}\sum_F A_L(f)A_S^*(f),
\label{bsr}
\end{eqnarray}
where $\phi_{SW} = \rm{arctan}[2(m_L-m_S)/(\Gamma_S-\Gamma_L)]$.  One can see that the Bell-Steinberger relation \eqref{bsr} relates a possible violation of CPT invariance ($m_{K^0}= m_{\overline{K}^0}$ and/or 
$\Gamma_{K^0}=\Gamma_{\overline{K}^0}$) in the $K^0-\overline{K}^0$ system to the observable CP-violating interference of $K_S$ and $K_L$ decays into the same final state $f$. If CPT invariance is not violated, 
then  $Im(\delta)=0$. We stress that  any evidence for $Im(\delta)\neq 0$ resulting from this relation can only manifest the violation of CPT or unitarity \cite{adpdg}. 

Generally, the advantage of the neutral kaon system is attributed to the fact, that only a few (hadronic)  decay modes give significant contributions to the rhs of Eq. (\ref{bsr}).
However,  what are the contributions from $\ksl$ decay modes and  how much the errors on 
$Re(\epsilon)$ and $Im(\delta)$ would  increase if these modes have maximal CP violation
are  still open questions, see, e.g.,  Ref.\cite{worksh}, that have to be  answered experimentally.
\par Using  the results of the most precise  measurements of the branching fractions of the visible 
$K_S,~ K_L$ decay modes from Particle Data Group PDG \cite{pdg} the estimate of the allowed extra contribution of $\kslinv$ decays to the 
total  decay rate of $K_S$ and $K_L$ result, respectively,   in
 \begin{equation}
 Br(K_S \to invisible) < 1.1 \times 10^{-4}, ~ ({\rm  ~95\%~ C.L.}),
  \label{eq:brks}
 \end{equation}
  and 
 \begin{equation}
 Br(K_L \to invisible) < 6.3 \times 10^{-4}, ~({\rm ~95\%~ C.L.}), 
 \label{eq:brkl}
 \end{equation}
 that would be interesting to check, see  Ref. \cite{Gninenko} for more discussions.

\section{Models with  $\ksl$  decays}
Being motivated by the above  considerations,  we discuss in this section several natural extensions of the SM predicting the existence of invisible   $K_S,~K_L$
decays \cite{GK2015, GK2016}. 
We show that taking into account the most stringent constraints from the measured $K^+ \rightarrow \pi^+  +invisible$ decay rate, the decay $\ksl$ could occur  at the level 
$Br(\ksl)\simeq 10^{-8} - 10^{-6}$. The main feature of the considered models,  that leads to the enhanced $Br(\ksl)$  compared to those from Eqs.(\ref{klsm},\ref{k+sm}), is that they allow to avoid the helicity suppression factor $\Bigl(\frac{m_\nu}{m_{K_L}}\Bigr)^2$ of the SM, while profiting from its larger phase-space due to the decay into two  light weakly interacting particles. In addition, there might be the case when $\ksl$ could still be kinematically allowed,  while  $K^+ \to \pi^+ + invisible$ is forbidden.
\subsection{The model with additional Higgs isodoublet}
Probably the simplest model predicting $K_S,~K_L$ invisible decays is the model with additional Higgs isodoublet
$H_2 = (H_2^+, ~H_2^{0})$  \cite{GK2015}.
  The additional Higgs isodoublet with zero vacuum expectation value $<H_2> = 0$ interacts with  quarks
  generations, namely
  \begin{equation}
        L_{H_2, ~quarks} =  -h_{ij}\bar{Q}_{Li}H_2 d_{Rj} + h.c. \,,
\label{YukH2}
  \end{equation}
  where $d_{R1} = d_{R}, ~d_{R2} = s_{R}, ~d_{R3} = b_{R}$ and $ Q_{L1} = (u,~d)_L, ~ Q_{L2} = (c,~s)_L ~ Q_{L3} = ~(t,~b)_L$.
  We  discuss the physics of $K$-mesons so we omit the effects related with the third generations of quarks.
  In general for $h_{12}h^*_{21} \neq 0$ the interaction (\ref{YukH2}) leads to flavour violating $\Delta S  = 2$ currents.   
  The measured $K_L ~-~K_S$ mass difference and the CP-violation parameter $\epsilon_K$ strongly restricts
  \cite{DeltaS=2} the effective $\Delta S = 2$ interaction
\begin{equation}
   L_{\Delta S=2} =  \frac{1}{\Lambda^2_{\Delta S =2}} \bar{d}_Ls_R \bar{d}_R s_L + h.c. \,,
    \label{DeltaS=2}
\end{equation}
namely \cite{DeltaS=2}
  \begin{equation}
    |Re(\Lambda_{\Delta S =2}| \geq 1.8 \cdot 10^{7}~GeV \,,
    \label{ReLambda}
  \end{equation}
\begin{equation}
    |Im(\Lambda_{\Delta S =2}| \geq 3.2 \cdot 10^{8}~GeV \,,
    \label{ImLambda}
\end{equation}
For the considered model (\ref{YukH2}) we find that 
\begin{equation}
   \frac{1}{\Lambda^2_{\Delta S =2}} = \frac{h_{12}h^*_{21}}{M^2_{H_2}} \,.
    \label{DeltaS=2h_12}
\end{equation}
We shall assume that $h_{12} $ and $h_{21}$ are real. As a consequence the Yukawa
      interaction (\ref{YukH2}) is CP-conserving and the most strongest bound (\ref{ImLambda})
       on the $\Delta S= 2$ currents is avoided.
   For CP-conserving interaction  as a consequence of the formulae
(\ref{ReLambda}) and      (\ref{DeltaS=2h_12})
  the bound on  the mass of the second Higgs isodoublet reads 
  \begin{equation}
    M_{H_2} \geq 1.8|h_{21}h_{12}|^{1/2}\cdot 10^{7}~GeV \,.
\label{H2mass}
  \end{equation}

  In considered model  the second  Higgs isodoublet $H_2$ interacts also with leptons 
  \begin{equation}
    L_{H_2,~leptons} = -h_{Lk}\bar{L}_k\tilde{H}_2\nu_{Rk} + h.c. \,,
    \label{nulept}
    \end{equation}
  where $L_1 = (\nu_{Le}, ~e_{L})$, $L_2 = (\nu_{L\mu}, ~\mu_{L})$, $L_3 = (\nu_{L\tau}, ~\tau_{L})$,
   $\tilde{H}_2 = (-(H^0_2)^*, ~(H^+_2)^*)$ and $\nu_{Rk}$ (k = 1,2,3) 
    are righthanded neutrino.
    We assume that the righthanded neutrino masses $m_{\nu_{Rk}}$
    are  much smaller the $K^0$-meson mass. As a consequence of the interactions
    (\ref{YukH2})
       and (\ref{nulept}) the $K_L,K_S$ mesons will decay invisibly into
  $K_L,K_S \rightarrow \nu_{Lk}\bar{\nu}_{Rk},~\nu_{Rk}\bar{\nu}_{Lk}$
       with the decay width
  \begin{equation}
    \Gamma(K_L(K_S) \rightarrow \nu_{Lk} \bar{\nu}_{Rk}, ~ \nu_{Rk} \bar{\nu}_{Lk}) =
    \frac{M^5_{K_L}}{16\pi M^4_X}(\frac{F_K}{2(m_d + m_s)})^2 K(\frac{m^2_{R1}}{M^2_{K_L}}) \,,
      \label{KLdecay}
  \end{equation}
  where
  \begin{equation}
    \frac{1}{M^4_X} = \frac{|(h_{12} +(-) h_{21})|^2  \cdot (|h_{L1}|^2   +  |h_{L2}|^2 + |h_{L3}|^2)   }{M^4_{H_2}} \,
      \label{KLdecay2}
  \end{equation}
  and $K(x) = (1-x)^2$ for Majorana neutrino with a mass $m_{R1}$ and massless neutrino $\nu_{L1}$
  \footnote{In formula (\ref{KLdecay2}) the sign $+$ corresponds to the $K_L$- decay and the sign $-$
    corresponds to the $K_S$-decay}.
  Here $F_K \approx 160~MeV$ is kaon lepton decay constant and $m_d$, $m_s$  are the masses of d-
  and s-quarks. In our estimates we take $(m_s +m_d)(2~GeV) = 100~MeV$ and
  $\Gamma_{tot}(K_L) =  1.29 \cdot 10^{-17}~GeV$, $\Gamma_{tot}(K_S) =  7.35 \cdot 10^{-15}~GeV$.
  Using the formula (\ref{KLdecay})  we find that for
  $Br(K_L \rightarrow \nu_{Ll} \bar{\nu}_{Rk}, ~ \nu_{Rk} \bar{\nu}_{Lk}) = 10^{-6}$ we can test the values of $M_X$
  up to 
  \begin{equation}
    M_X \leq 0.74\cdot10^{5}~GeV(    1.5\cdot10^{4}~GeV )    \,.
    \label{MXboundKL}
  \end{equation}
  for $K_L(K_S)$-mesons.

  It should be noted that the bound  (\ref{H2mass}) strongly restricts but not excludes
  phenomenologically interesting values of $M_X$
and invisible neutral $K$-meson decays with the branching at the level of $O(10^{-6})$.
  For instance, for $h_{12} = h_{21} = 2\cdot 10^{-5}(2\cdot 10^{-4})$,   $h_{L1} = h_{L2} = h_{L3} = 1$ and $M_{H_2} = 400(4000)~GeV$
  we find that $\Lambda_{\Delta S = 2} = 2   \cdot 10^{7}~GeV$ and
  $Br(K_L \rightarrow \nu_k \bar{\nu}_k)=   5.7  \cdot 10^{-6}(5.7 \cdot 10^{-8}) $.
  For the case $h_{12} = 0$ or $h_{21} = 0$ the bound (\ref{H2mass}) dissapears.

 \subsection{The model with additional scalar isodoublet and isosinglet}
    In this subsection we consider the modification of the previous model. Namely in addition to the
    second Higgs isodoublet $H_2$ we add neutral scalar isodoublet $\phi$ with the interaction
    \begin{equation}
      L_{H_2 H \phi \phi} = -\lambda H^+_2 H \phi \phi + h.c. \,.
      \label{H2Hphiphi}
    \end{equation}
    After electroweak symmetry breaking the effective trilinear term
    \begin{equation}
      L_{H_2 \phi \phi} =
      -\lambda <H> (H_2^0  + (H_2^0)^*) \phi \phi \,
      \label{H-2phiphi}
    \end{equation}
becomes  responsible for the interaction of $\phi$-particles with quarks. 
     Here $<H> = 176~GeV$. The effective Lagrangian
    \begin{equation}
      L_{eff} =     \frac{\lambda <H>}{M^2_{H_2}}        [h_{12}\bar{d}_Ls_R  + h_{21} \bar{s}_Ld_R  + h.c. ]  \phi^2  \,.
      \label{Leff}
    \end{equation}
      The $K_L$ decay width $K_L       \rightarrow \phi \phi $ has the form
      \begin{equation}
    \Gamma(K_L \rightarrow   \phi \phi ) =      (\frac{\lambda <H>}{M^2_{H_2}})^2   |h_{12} + h_{21} -h^*_{12}- h^*_{21}|^2
    \frac{M^3_{K_L}}{64\pi }(\frac{F_K}{2(m_d + m_s)})^2 K(\frac{m^2_{\phi}}{M^2_{K_L}}) \,,
      \label{KLdecay2a}
  \end{equation}
     where $K(x) =  (1-4x)^{1/2}$.
     The formula for $K_S$ decay width has the similar form
  \begin{equation}
    \Gamma(K_L \rightarrow   \phi \phi ) =      (\frac{\lambda <H>}{M^2_{H_2}})^2              |h_{12} - h_{21}  + h^*_{12}- h^*_{21}|^2
    \frac{M^3_{K_L}}{64\pi}(\frac{F_K}{2(m_d + m_s)})^2 K(\frac{m^2_{\phi}}{M^2_{K_L}}) \,,
      \label{KLdecay2b}
  \end{equation}


Note also that the model proposed in ref.\cite{GK2016} also predicts 
the existence of  $K_L, K_S$ invisible decays. The peculiarity of the model
\cite{GK2016}  is the use of nonrenormalizable Lagrangian
  \begin{equation}
        L_{\phi, ~quarks} =  -h_{ij}\bar{Q}_{Li}H q_{Rj}\phi + h.c. \,.
        \label{YukHphi}
        \end{equation}
Here,  $\phi$ is a neutral $(\phi^* = \phi)$  scalar field with a mass $m_{\phi}$ and $ H = (H^+, H^0)$ is the SM Higgs isodoublet.  
It is assumed also  that the field $\phi$ couples  with righthanded neutrino (dark matter). 
 The corresponding formulae for $K_L, K_S$ invisible decay widths  and the
  bounds on the $h_{ij}$ coupling constants are contained in ref.\cite{GK2016} and they are similar to the formulae (\ref{H2mass},\ref{KLdecay}).

\section{Oscillations of $K_S,~K_L$ into the dark mirror sector}
Today the origin of dark matter (DM) in the Universe is still a great puzzle, see, e.g. \cite{Rubakov:2017xzr, Kolb:1990vq}.
  The possible existence of the dark mirror sector part of which is mirror DM with particle and interaction content identical to  a mirror copy of the SM 
  has still received significant attention, as the DM  could be explained by the existing of the mirror baryons  \cite{mirror, mirrormatter, footdm, wil}.
   If the  dark mirror sector exists, the mixing between the SM neutral states,  such as 
positronium ($Ps$)- mirror-positronium ($Ps^M$) \cite{pos}, neutron ($n$)-mirror-neutron ($n^M$) \cite{zurab}, and $K^0$ - mirror-$K^{0M}$ \cite{okunnik} are possible
resulting in oscillations of SM states into the corresponding mirror ons.  
The  searches for this effect have been performed for the $Ps-Ps^M$ \cite{paolo} and $n-n^M$ \cite{psinn1,psinn2,ser} oscillations, but not for the $K^0-K^{0M}$ system yet. 
\par Probably,  the simplest model predicting the $K^0 - K^{0M}$ oscillations is the following \cite{GK2015}.
In the SM and  mirror SM models two additional Higgs isodoublets $H_2$ and $H_2^M$ are introduced and the interaction of
$H_2(H^M_2)$ isodoublets with quarks(mirror quarks) has the form \ref{YukH2}). The interaction of $H_2$, $H_2^M$ isodoublets with
the standard $H$ and $H^M$ isodoublets has the form
\begin{equation}
  L_{H_2H_2^M} = - \lambda_{M} (H_2^+H)(H^+H^M_2) + h.c.
  \label{mirrorint}
  \end{equation}
After electroweak symmetry breaking  we find that the effective interaction
\begin{equation}
  L_{eff,mix} = - \lambda_{M}(<H>)^2 (H_2^{0})^*(H^{0M}_2) + h.c.
  \label{mirroreff}
  \end{equation}
is responsible for  the  mixing of $K^0$ and  $K^{0M}$ mesons.
Namely, as a consequence of nonzero mixing (\ref{mirroreff}) the four-fermion interaction
\begin{equation}
L_{4ferm,mix} = \frac{1}{M^2_{eff}}[(h_{ij}\bar{d}_{Li}d_{Rj})\cdot (h_{ij}\bar{d}^M_{Li}d^M_{Rj})^{*} +  
    [(h_{ij}\bar{d}_{Li}d_{Rj})^{*}\cdot (h_{ij}\bar{d}^M_{Li}d^M_{Rj})] \,,
    \label{mirrorferm}
\end{equation}
leads to the oscillations between $K^0$ and $K^{0M}$ mesons. Here
\begin{equation}
  \frac{1}{M^2_{eff}} =    \frac{\lambda_M<H>^2}{M^4_{H_2}} \,.
  \label{Meff}
  \end{equation}
 

\par As an example consider  the case of $K_L - K^M_L$ oscillations.
The mixing between $K_L$ and  $K_L^M$ mesons is described by the effective Hamiltonian
\begin{equation}
  H_{mix} = \delta K_LK^M_L \,,
  \label{mirrormix}
\end{equation}
The states $K_{\pm} = \frac{1}{\sqrt{2}}(K_L \pm K^M_L)$  have the masses
$m_{\pm} = (m_{K_L} \pm \frac{\delta}{2})$.  An ordinary $K_L$  produced in strong interactions
would oscillate into mirror  $ K_L^M$ state with  the probability determined by 
\begin{equation}
  P(K_L \rightarrow K^M_L|t)  =    |<K^M_L(t)|K_L(t=0)>|^2   = \frac{1}{4}|1 - \exp(-i\delta t)|^2 exp(-\Gamma_{K_L} t).             
\label{oscillation}
\end{equation}
The full probability at the time    $t_0 \geq t \geq 0$  is given by the formula
\begin{equation}
 P_{int}(K_{L} \rightarrow K_L^M| t \leq t_0)  \equiv
 \int^{t_0}_{0}dt  |<K^M_L(t)|K_L(t=0)|>^2    \,,
 \label{probt0}
\end{equation}
resulting in 
\begin{equation}
\begin{split}
&P_{int}(K_{L} \rightarrow K_L^M| t \leq t_0)  =   \frac{\delta^2}{2\Gamma_{K_L}(\Gamma^2_{K_L} + \delta^2)} \cdot \\
&(1  - \exp(-\Gamma_{K_L}t_0)) +  \frac{\Gamma_{K_L}(\cos(\delta t_0) - 1)- \Delta \sin(\delta t_0)}{2 (\Gamma^2_{K_L} + \delta^2) } \exp(-\Gamma_{K_L}t_0)).
\end{split}
\label{fullprob11}
\end{equation}
The integration over time gives the full probability for  the $K_{L} \rightarrow K_L^M$ conversion:
\begin{equation}
  P_{int}(K_{L} \rightarrow K_L^M|t\leq \infty) =
  \frac{\delta^2}{2\Gamma_{K_L}(\Gamma^2_{K_L} + \delta^2)}
  \label{fullprob1}
  \end{equation}
The relative probability for $K_L$  to oscillate into $K_{L}^M$ state at the time    $t_0 \geq t \geq 0$ is determined by the formula
\begin{equation}
  P_{rel}(K_L \rightarrow K_L^M| t\leq t_0) \equiv
  \frac{\int^{t_0}_{0}dt  |<K^M_L(t)|K_L(t=0)>|^2}{ \int^{t_0}_{0}dt  |<K_L(t)|K_L(t=0)>|^2}    
    \label{fullprob12}
  \end{equation}
In particular 
\begin{equation}
  P_{rel}(K_L \rightarrow K_L^M| t\leq \infty) =
  \frac{\delta^2}{2(\Gamma^2_{K_L} + \delta^2)}
\label{fullprob2}
  \end{equation}
  assuming $\Gamma_{K_L} \gg  \delta $.

Note that all previous formulae were derived for $K$-mesons in the rest frame. For kaons moving with the momentum $\vec{p}$
we must have the replacements $ m_{\pm}  = m \pm \frac{\delta}{2} \rightarrow  \tilde{m}_{\pm} =    m   \pm \frac{\delta}{2} - i \frac{\Gamma_{K_L}}{2}$
    in formulae $E_{\pm} = \sqrt{\vec{p}^2 + m_{\pm}^2}$ for kaon energies. For $ m \gg \delta, ~\Gamma$  we have to  perform  the replacements
    $\Gamma \rightarrow \Gamma \cdot \frac{m}{\sqrt{\vec{p}^2   + m^2}}$,  $\delta \rightarrow \delta \cdot \frac{m}{\sqrt{\vec{p}^2   + m^2}}$
      in formulae  (\ref{oscillation}, \ref{fullprob1}), which take the form
\begin{equation}
\begin{split}
  P(K_L \rightarrow K^M_L|t) =
\frac{1}{2}(1 - \cos (\delta \frac{m}{\sqrt{\vec{p}^2   + m^2}}t))  exp(-\Gamma_{K_L} \cdot  \frac{m}{\sqrt{\vec{p}^2   + m^2}} t), 
  \label{oscillation2}
\end{split}
\end{equation}
and 
\begin{equation}
  P_{int}(K_{L} \rightarrow K_L^M)|t \leq \infty)    =
  \frac{\delta^2}{2\Gamma_{K_L}(\Gamma^2_{K_L} + \delta^2)} \cdot \frac{\vec{p}^2 + m^2}{m^2} \,,
  \label{fullprob2}
  \end{equation}
  respectively.

\section{A search for  $\ksl$ decays and oscillations into dark sector} 
\label{sec:ExpInvisible}
In this section we discuss briefly  an experiment on searching for $\ksl$ decays and $K_{S,L} - K_{S,L}^M$ oscillations within the 
NA64 experimental program at CERN, see, e.g. \cite{Gninenko:2020hbd, Gninenko:2021}.
The signature of the latter would be the disappearance of the $K_{S,L}$ from the beam due to their $\ksl$ decays in flight into invisible dark final states. 
Searching  for $\ksl$ decays is challenging, as it requires a combination of  an intense source of  $K^0$s and a well-defined high-purity signature 
to tag their production. Currently there is no experimental limits on $\kslinv$ decay modes, apart from those, see \eqref{eq:brks} and \eqref{eq:brkl},  obtained assuming  that unitarity is the fundamental property of the $K_{S,L}$ decays \cite{Gninenko}.

\subsection{The search method and the experimental setup}

The general method for searching for neutral meson $M^0 (\pi^0, \eta, \eta', K_S,K_L..) \to invisible $ decays was proposed in Ref. \cite{Gninenko}. 
Recently, the NA64 collaboration at CERN obtained the first proof-of-concept 
results on the search for  $\eta, \eta' \to invisible$ decays based on this method \cite{na642024}, which  is briefly described below. 
\par The source of  $M^0$ could be  either the quasi-elastic charge-exchange reaction of high-energy kaons on nuclei of an  active target 
\begin{eqnarray}
 K^- + A(Z) \to \overline{K}^0 + n + A(Z-1),& ~\text{or}   \nonumber \\ 
K^+ + A(Z) \to K^0 +p+A(Z)
\label{eq:kchex}
\end{eqnarray}
 or,  high-energy $\pi^{\pm}$ induced reactions
\begin{equation}
\begin{split}
&\pi^- + A(Z) \to \eta, \eta', ... + n +A(Z-1) \\   
&\pi^- + A(Z) \to K^0 + \Lambda +A(Z-1) \\
&\pi^+ + A(Z) \to K^0 + \Sigma^+ +A(Z)  
\end{split}
\label{eq:pichex}
\end{equation}
The reactions with $\pi^\pm$  is more interesting for the $\ksl$ search compared to  \eqref{eq:kchex}, due to the significantly higher intensity of $\pi^\pm$ beams and comparable cross sections of the $K^0$ production in  \eqref{eq:kchex} and \eqref{eq:pichex}.  In these  quasi-elastic  processes, e.g.  the neutral kaon is emitted mainly in the forward direction with the beam momentum and the recoil nucleon/nuclei carries away a small fraction of the beam energy, 
The term "quasi-elastic reaction"  means that, unlike elastic reactions of charge exchange with the proton(neutron), the transition can occur for the target  nucleus as a whole into an excited  state followed by its fragmentation. Since the binding energy in the nucleus is a few MeV/nucleon, the velocity $v\sim q/$mass of the daughter particles, where $q\lesssim0.05$ GeV/c is the momentum transfer, is on average small. At high initial energies, the nucleus does not have time to collapse during the interaction (the characteristic transverse distances is $l\simeq 1 /q$). After the collision, the nucleus disintegrates into fragments, which are absorbed into the target. Hence the experimental signature of the $M^0$ production in  \eqref{eq:kchex} or \eqref{eq:pichex}, is an event with  {\it  full disappearance of the beam energy}. The  decay $\kslinv$ is expected to be a very rare event that  occurs with a much smaller frequency than the  $\ksl$ production rate. Hence, its observation presents a challenge for the design and performance of the detector. However,  despite a relatively  small $\ksl$ production rate  the  signature of the signal event is very powerful  allowing  a strong background rejection. 

\begin{figure*}[tbh!]
\includegraphics[width=.95\textwidth]{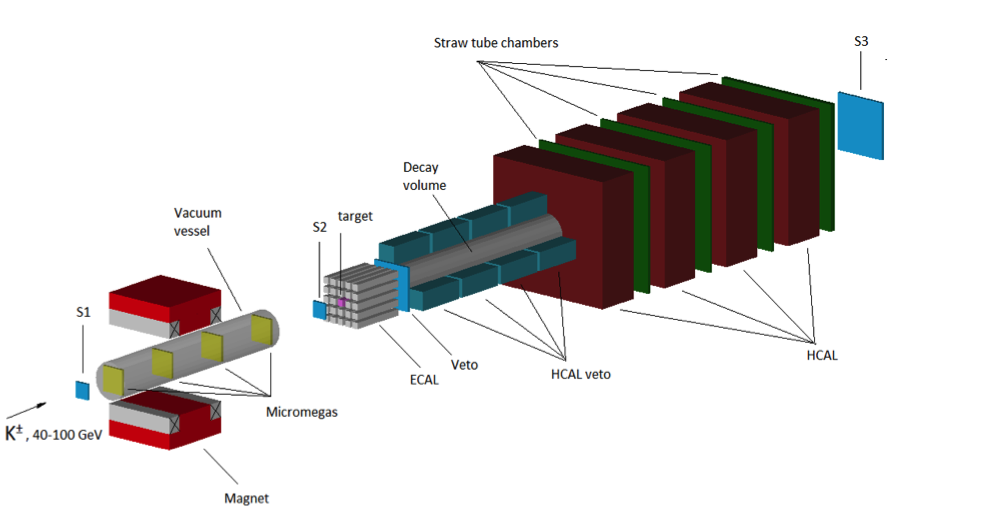}
\caption{\label{fig:setup} 
Schematic illustration of the setup to search for the $M^0\to invisible$ decays and $K_{S,L} - K_{S,L}^M$ oscillations  (see text). }
\label{setup}
\end{figure*}

 The  detector designed to search for the  $\kslinv$ decays is schematically shown in 
Fig. \ref{setup}, and  is complementary to the one  proposed for the NA64 search for
invisible decays of dark photons at  CERN \cite{sngldms, ldms}.  It is also  similar to the setup used by NA64 to search for $\eta,\eta' \to invisible$ decay modes \cite{na642024}. 
The major distinction is  adding a new segmented target for the reactions    \eqref{eq:kchex}, \eqref{eq:pichex} and an additional  vacuum  decay volume surrounded by a veto system to look for $K_{S,L}$ oscillations to the dark sector, see below.
The experiment  employs, the T9 (or H4)  $\pi^\pm$ and $K^\pm$  beams, which are  produced in a target of the CERN PS (SPS)  and transported to the detector by a beamline tuned to a freely adjustable  beam momentum  around$\sim$ 15 GeV/c \cite{sps}. 
The maximal T9 beam  intensity  is $\simeq10^6~ \pi^\pm$ with the  fraction of $ K^\pm$ $\sim$ a few \% per PS spill. The typical PS cycle for 
fixed-target (FT) operation lasts 16.8 s, including  from one to five  spills of $\sim$ 0.4 s duration depending on the accelerator regime.
 The maximal number of FT cycles is four per minute.  The 
beam has low purity - the   admixture of the other charged particles is  a few \%. It  can be focused onto a spot of the order of a few cm$^2$. 
The incident charged particle is defined by the scintillating counters S1,S2. The momentum of the beam is additionally selected  with a momentum spectrometer
consisting of a dipole magnet and a low-density tracker, made of a set of  Micromegas detectors (MM)  or Straw Tube chambers (ST). 
The setup is a completely hermetic detector allowing to measure accurately the full energy 
 deposition from the reactions \eqref{eq:kchex}, \eqref{eq:pichex}.  It  is  equipped with an active target $T$ made of a segmented scintillator counters,   surrounded by a  high-efficiency  electromagnetic calorimeter (ECAL) serving  as a veto against photons and other secondaries emitted from the target  at large angles and mimic the reactions  \eqref{eq:kchex}, \eqref{eq:pichex}, high-efficiency forward veto counter Veto, a decay vacuum volume DV surrounded by a thick veto hadronic  calorimeter modules (HCAL veto),  followed by  a massive, hermetic hadronic calorimeter (HCAL) located at the end of the setup and separated by a large size Straw Tube chambers. 
 For searches at low  energies,  Cherenkov counters  to tag  the incoming hadron and enhance it identification (ID)  can be used.

 The reactions~  \eqref{eq:kchex}, \eqref{eq:pichex} occurs practically uniformly over the target  length. 
 The distribution  of the primary kaon (pion) energy deposited in the target can be used  as a signature of the $M^0$ production, see Sec.\ref{sec:tag}, and to determine the position
of the interaction vertex along the beam direction.
 The produced   $K^0$ - composed of equal portions of $K_S$ and $K_L$-  either 
decay quickly in the target $T$, or  penetrates the veto system without interactions  and either decays in flight  in the DV  or interacts in the HCAL. If the $K_S$ and $K_L$ decay invisibly, it is assumed that  the final-state  particles in this case also penetrate the rest of the detector without prompt decay  
into ordinary particles, which could deposit energy in the HCAL. The full HCAL calorimetric system surrounding the decay volume, 
is designed  to detect with high efficiency the energy deposited by secondaries from the  primary  interactions $K^\pm A \to anything$  in the  target. 
  In order to suppress  background due to the detection inefficiency,  the detector must be longitudinally completely hermetic. To enhance detector hermeticity, the HCAL  calorimeter has a total thickness of    $\simeq 28 ~\lambda_{int}$ (nuclear interaction lengths) and has to have high light-yield to minimize background from the statistical fluctuations of the photoelectrons. 

\par The setup configuration shown in Fig.\ref{setup} also allows to search for $K_{S,L} - K^M_{S,L}$ oscillations.
An ordinary $K_L$  produced in strong interactions, e.g. of \eqref{eq:pichex}, 
would oscillate into mirror  $ K_{S,L}^M$ state with  the probability determined by the Eq.(\ref{oscillation2}). 
The occurrence of the  $K_{S,L} -  K_{S,L}^M$ transition would exhibit itself as the disappearance of the $K_{S,L}$'s 
 from the beam,  i.e. as the $\ksl$ decay, with the rate given by Eq. (\ref{fullprob2}) as a function of the $K_{S,L}$ flight-time $t$, assuming $\tau_S < t < \tau_L$.  
The occurrence of $\kslinv$ decays produced in $K^\pm$ interactions would appear as an excess of events with a signal in the $T$, see Fig.~\ref{setup} and zero energy deposition in the rest of the detector (i.e. above that expected from the background sources). 

\subsection{The $K^0$ tagging system}
\label{sec:tag}
\begin{figure*}[tbh!]
\centering
\includegraphics[width=.8\textwidth]{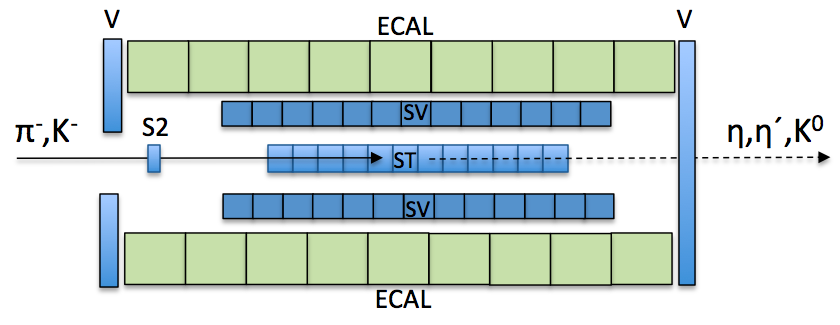}
\caption{\label{fig:setup} 
Schematic illustration of the target equipped to search for the invisible decays of neutral 
kaons in the proposed  experiment and tagging the  reaction \eqref{eq:kchex}.
The incoming $K^-$ is defined by the counter S, SV is the scintillator veto-counters, ECAL is a guard veto electromagnetic calorimeter 
 against the electromagnetic secondaries.
}
\label{fig:target-K0}
\end{figure*}

To reduce the counting rate and ensure the effective $K^0$  selection, one could use a system surrounding  the T target for the  efficient 
tagging  of the $K^0$ production. 
 For reactions \eqref{eq:kchex}, \eqref{eq:pichex} the schematic illustration of the $K^0$ tagging system is shown in Fig.\ref{fig:target-K0}. 
For example, the incoming $\pi^-$ or $K^-$ defined by the scintillator counter $S2$ enter a segmented  target (ST), which consists of 12 scintillator cells 
numbered from $i=1$ to 12, and  produce the leading $K^0$ accompanied by a low energy recoil neutron or nuclear fragments. 

The occurrence of $\kslinv$ decays produced in $K^-$ interactions would appear as an event with a signal in the $T$, see Fig.~\ref{fig:target-K0} and zero energy deposition in the rest of the detector. Thus, the signal candidate events have the signature 
\begin{equation}
S_{\koinv} = {\rm T \cdot \overline{SV\cdot ECAL\cdot V\cdot HCAL}}
\label{eq:sign} 
\end{equation}
and should satisfy the  following selection criteria:
\begin{enumerate}[(i)]
\item The measured momentum of the incoming kaon should correspond to its selected value.
\item The kaon should enter the target and the  interaction vertex  should be localized  within the target cell $T_j$  with $1 < j < 12$, with a MIP signal in cells 
with the number $i<j$, and no signal in cells $i>J$.
\item The should be no energy deposition in  the ECAL veto, SV and V.   
\item The fraction of the beam energy deposited in the veto HCAL modules and HCAL should be consistent with zero.
\end{enumerate}
 
In the case of using  the reaction \eqref{eq:pichex} as the source of $K^0$'s one can additionally tag them via  the presence of the $\Lambda$ decay products.
The  segmented SV counters surrounding the target $T$ can be used  to detect $\pi^- + p$ pairs from the decay $\Lambda \to \pi^- + p$, while  
the $\g \g$ pair from the decay chain $\Lambda \to \pi^0 + n; \pi^0 \to \g \g$ could be registered in the segmented ECAL calorimeter surrounding the $T$, 
 as shown in Fig. \ref{fig:target-K0}.
The development of  such tagging  system, including possible detection of recoil neutrons from the reaction \eqref{eq:kchex}, is currently under consideration.     

  
\subsection{Background and expected sensitivity}
The background processes  resulting in  the signature of the primary reaction \eqref{eq:kchex}, and similar of \eqref{eq:pichex}, 
  can be classified as being due to physical-  and  beam-related sources.  To perform a full detector simulation in order  to investigate these backgrounds down to the level  $ \lesssim 10^{-10}$  would require  a prohibitively large amount of computer time. Consequently, only the following sources of background - identified as the most dangerous - are considered and evaluated  with  reasonable statistics combined  with numerical calculations:

\begin{enumerate}[(i)]
\item One of the main background sources is related to the  low-energy tail in the distribution of the energy of the primary  hadronic beam. This tail is caused by the beam interactions with a passive material, such as the entrance windows of the beam lines, residual gas, etc. Another source of low-energy hadrons  is due to beam $\pi^\pm, K^\pm$ decays in flight into a low-energy secondary electron,  pions or muons that mimic the signature \eqref{eq:sign}  in the detector.  For example, the beam $\pi^-$ or $K^-$ meson could decay into 
a backward $e^-$ or $\mu^-$ with a very low energy,  $\lesssim 50$ MeV,  that stop in the target mimicking the charge-exchange signature.  
To improve the primary high-energy hadron selection  and suppress background from the possible admixture of 
low-energy particles, one can use a tagging system utilizing  the magnetic spectrometer installed upstream of the detector,  shown in Fig.~\ref{setup}, and  the $K^0$ tagging system discussed in Sec.\ref{sec:tag}. Additionally, Cherenkov counters can be used to identify  kaons which are expected to be the main source of this background. 
\item
The background events  could also arise when the leading $K_L$ or neutron  from the reaction $\pi+A \to K_L,~n + X$ that occurred in the target is not detected due to the incomplete hermeticity of the HCAL. In this case, e.g. the $K_L$ punches through the HCAL without depositing energy above a certain threshold $E_{th}$. 
The punch-through  probability is defined roughly by $\simeq exp(-L_{HCAL}/\lambda_{int})$,
where $L_{HCAL}$ is the HCAL thickness. Thus, by selecting the total HCAL thickness about 28 $\lambda_{int}$ this background can be suppressed 
down to the level  $ \simeq 10^{-12}$. 
 \item Another type of background process  is caused by $\pi,K\to \mu,e + \nu$  decays  in-flight  of pions and kaons  after they have passed the magnetic spectrometer.
 The background of the low-energy  muon admixture  in the beam from the $\pi,K\to \mu \nu$  decays
  can be due to the following event chain. The decay muon entering the detector decays in flight into  a low-energy electron and a neutrino pair, $\mu \to e \nu \nu$ in the target. The electron then penetrates Veto without being efficiently detected, and deposits all its energy  in the HCAL, 
which is below the threshold $E_{th}\lesssim 0.5$ GeV. The probability for this  event chain  is found to be as small
as $P\lesssim 10^{-12}-10^{-11}$. Similar background caused by the decays of the beam pions or kaons in the target was also found to be negligible. 

\item
The fake signature  could be due to the physical background: a muon scattering on a nucleon, e.g. 
  $\mu^- p \to \nu_\mu n$, accompanied by a poorly  detected neutron. Taking into account the corresponding cross section and the probability for the 
  recoil neutron to escape detection in the HCAL results in an overall level of this background 
  of $\lesssim 10^{-12}$ per incoming hadron.
\end{enumerate}

In Table~\ref{tab:table1} contributions from the all background processes are summarized for the primary 
$\pi^-$ beams with energy 15 GeV. The total background is expected  to be at the level $\lesssim 10^{-11}$ per incoming  pion. 
Therefore, the search accumulated up to a few $10^{11}$ events is expected to be background free. The expected sensitivity in branching fractions is summarized below.
assuming the background free search.  
\begin{table}[tbh!] 
\begin{center}
\caption{Expected contributions to the total level of background from different background sources estimated 
per incident  $\pi^-$,  (see text for details).}\label{tab:table1}
\vspace{0.15cm}
\begin{tabular}{lr}
\hline
\hline
Source of background& Expected level\\
\hline
HCAL nonhermeticity & $ \lesssim   10^{-12}$\\
punch-through $K^0$s,    &$ \sim 10^{-12}$\\
$\pi^-, K^-\to \mu^- \nu+X $ decays in flight & $ \lesssim 10^{-11}$\\
$\pi^-,K^- \to e^- \nu+X$ decays in flight & $\lesssim 10^{-12}$\\
$\mu^-$ induced  reactions on target nuclei & $\lesssim 10^{-12}$\\
very low-energy tail of the beam& $ \lesssim  10^{-11}$\\
\hline 
Total (conservative)  &   $ \lesssim  10^{-11}$ per incoming \\
\hline
\hline
\end{tabular}
\end{center}
\end{table}

To estimate the sensitivity of the proposed experiment 
 a simplified feasibility study  based on GEANT4 \cite{geant}
Monte Carlo simulations have been  performed for 15 GeV pions and kaons. The ECAL is an  array of  the lead-scintillator  counters 
allowing for accurate measurements of the lateral energy leak from the target. The target is a set of plastic scintillator cells with  thickness $\simeq 0.04\lambda_{int}$ viewed by a SiPM photodetector.  
The SV veto counters are of 1-2 cm thick, high-sensitivity Sc arrays with a high light yield  of $\simeq 10^3$ photoelectrons per 1 MeV of deposited energy. It is also assumed that the veto's inefficiency  for the MIP  detection  is, conservatively, $\lesssim 10^{-4}$. The hadronic calorimeter is a set of four modules. 
Each module is a sandwich of alternating layers of iron and scintillator with thicknesses of 25 mm and 4 mm,  
respectively, and with a lateral size of $60\times 60$ cm$^2$.    Each module consists of 48 such layers and has 
a total thickness of $\simeq 7~\lambda_{int}$. 
The number of photoelectrons produced by a MIP crossing the module  is in the range $\simeq$ 200-300 ph.e..
The probability for an event with the MIP energy deposited in the HCAL to mimic the signal  due to fluctuations of $n_{ph.e.}$ is negligible. 
 The hadronic energy resolution of the HCAL calorimeters as a function of the beam energy is taken to be 
$\frac{\sigma}{E} \simeq \frac{ 60 \%}{\sqrt{E}}$ \cite{ihephcal}. The energy threshold for the zero-energy in the HCAL is 0.1 GeV. The reported further analysis also  takes into account passive materials from  the DV vessel  walls.
The expected sensitivities for the  process  \eqref{eq:pichex} was estimated from 
the  calculation of  fluxes and energy distributions  of mesons produced in the target by taking into account  
the relative normalization of the yield of  $K^0$ from the original publications \cite{foley, yud1, yud2}. 
For the purpose  of this work,  the total  $K^0$  production cross sections in 
the $\pi, K^-$ charge-exchange reactions in the target  were calculated  from their  extrapolation to the target atomic number as described in 
Ref.\cite{aszh2}.  Note, that the yield of 
 $K^0$ is also supposed to be measured  {\it in situ} (see discussion below). Typically, the branching fractions of the charge-exchange reactions  are in the range
$\frac{\sigma(K^-p\to \overline{K}^0 n)}{\sigma(K^-p\to all)}\simeq \frac{\sigma(\pi^-p\to \pi^0 n)}{\sigma(\pi^-p\to all)}\simeq 10^{-4}-10^{-3}$ and depend on the beam energy \cite{foley, yud1, yud2}.

The calculated fluxes and energy distributions  of mesons produced in the target are used to predict the number of signal events in the detector. 
For a given number of primary pions $N_{\pi^-}$, the expected total number of 
$\ksl$ decays occurring within the decay length $L$ of the detector is given by 
\begin{equation}
n^{inv}_K = n^{inv}_{K_S} + n^{inv}_{K_L}
\label{ntot}
\end{equation}
with 
\begin{eqnarray}
&n^{inv}_{K_{S,L}}= k N_{\pi^-}Br(\ksl) \cdot \int\frac{\sigma(\pi^- +A \to K^0+.. )}{dt} \Bigl[1-{\rm exp}\Bigl(-\frac{L M_{K^0}}{P_{K^0}\tau_{K_{S,L}}}\Bigr)\Bigr]\zeta \epsilon_{tag}  dt \nonumber \\
& \simeq  \zeta \epsilon_{tag} Br(\ksl) n^{dec}_{K_{S,L}}
\label{nev}
\end{eqnarray}
where coefficient $k$ is a normalization factor that was
tuned to obtain the total  cross section of the meson production,  $  P_{K^0}$ and $\tau_{K^0}$ are the  $K^0$ momentum and the lifetime of either $K_S$ or $K_L$ at rest, respectively, $\zeta$ is the signal reconstruction efficiency, $\epsilon_{tag}$ is the tagging efficiency of the final state,  and $n^{dec}_{K_{S,L}}$ is the total number of $K_{S,L}$ decays occurring  in the decay volume of length $L$. In this estimate we neglect the $K^0$ interactions in the target: the  average momentum of the incoming kaons is in the range $<p_{K^-}>\simeq 15$ GeV, the decay length   $L\simeq 5$ m, and the efficiency $\zeta \simeq 0.9$. The tagging efficiency $\epsilon_{tag}$ is  typically $\gtrsim 90\%$ \cite{foley, yud1,yud2,gams}.
  
\par In the case of no signal observation, the obtained results can be used to impose upper limits  on the decays of $K_{S,L}$ into invisible final states; by using the relation $n^{inv}_K = n^{inv}_{K_S} + n^{inv}_{K_L}< n^{inv}_{90\%} $, where $n^{inv}_{90\%}$ (= 2.3 events) is the 90$\%$ C.L. upper limit for the  number of signal events,  and Eq. (\ref{nev}), one can  determine the expected $90\%~ C.L.$ upper limits from the results of the proposed experiment summarized in Table \ref{tab:lim} for the total number of $3\times 10^{11}$
pions on target. Here we also assume that the exposure to the $\pi/K$ beam with the nominal rate is a few months,
 and  that the invisible final states do not decay promptly into the ordinary particles, which would deposit energy in the veto system or HCAL.

 
\begin{table}
\caption{\label{tab:table2} Expected upper limits on the branching ratios of 
different decays into invisible final states calculated for the total number of $3\times 10^{11}$ incident 
pions and reaction \eqref{eq:pichex} as the source of $K_{S,L}$ ( see text for details). }
\begin{tabular}{lr}
Expected limits  &  Present limit\\
\hline
Br$(\ksinv) \lesssim  10^{-7}$ & no~~~~~~~ \\
\hline
Br$(\klinv) \lesssim  10^{-5}$ & no~~~~~~~  \\
\hline
Br$(\eta\to invisible )\lesssim 2.7 \times10^{-7}$ & $ < 1.0 \times10^{-4}$ \cite{bes14,na642024} \\
\hline
Br$(\eta' \to invisible )\lesssim 5.6\times 10^{-7}$ & $ < 2.1 \times 10^{-4}$\cite{na642024} \\
\hline
\hline
\end{tabular}
\label{tab:lim}
\end{table}
 
\par  Taking Eqs.(\ref{oscillation2}),(\ref{fullprob2}) into account the expected limits of Tab.\ref{tab:lim} can be transformed in the bound on the probability of the $ \kl - \klm$ oscillation and the 
 mixing strength $\delta$. 
For numerical estimates we shall take the energy of kaons equal to $E_K = 15~GeV$ and the decay length of
the detector for $K_S$, $K_L$ decays  equal to $L_0 = 5~m$. For $K_S$ mesons with the energy $E_0$ the decay length
$l_{K_S} = c\tau_{K_S}* \frac{E_K}{m_K} =  0.8 ~m  \ll L_0$. Here $c = 3\cdot10^{8} \frac{m}{sec}$ is the
velocity of light. It means that we can use the formula (\ref{fullprob2}) for the estimate of the
transition probability  $P(K_S \rightarrow K_S^M)$ of the $K_S$-meson into the dark $K_S$-meson.
Taking the bound on the invisible $K_S$-decay, $Br(K_S \rightarrow invisible) = 10^{-7}$ and using   Eq.(\ref{fullprob2}) we can obtain bound :
\begin{equation}
\frac{\delta_{K_S}}{\Gamma_{K_S}}  \leq 4.5 \cdot10^{-4}
\label{eq:limdelta}
\end{equation}
\par For $K_L$ mesons the situation is different. Indeed, for $K_L$-mesons the decay length
$l_{K_L} = c\tau_{K_L}* \frac{E_K}{m_K} =  466 ~m  \gg L_0$. It means that only a small fraction of the produced $K_L$s  decay inside the detector. 
As a consequence,  Eq.(\ref{fullprob11}) takes the form
\begin{equation}
  P_{int}(K_{L} \rightarrow K_L^M|t \leq t_o) \approx
  \frac{m^2_K}{12E^2_K}\delta^2 t_0^3 \,,
 \label{fullprob1a}
  \end{equation}
where $t_o \approx \frac{L_o}{c}$. The use  of the formula (\ref{fullprob1a}) gives 
\begin{equation}
  P_{rel}(K_L \rightarrow K_L^M) \approx
   \frac{m^2_K}{12E_K^2}\delta^2 t_0^2 \,, 
  \label{fullprob2ab}
  \end{equation}
Taking the bound  $Br(K_L  \rightarrow invisible) = 10^{-5} $ from Tab.\ref{tab:lim} and Eq. (\ref{fullprob2ab}) into account, one  can obtain for $K_L$ the  bound 
\begin{equation}
\frac{\delta_{K_L}}{\Gamma_{K_L}}  \leq 1, 
\end{equation}
 which is, as discussed above, rather weak due to the  fact that the fraction  of  $K_L$'s decaying at the length $0 \leq l \leq 5~m$ is just 
$1 - \exp(-\Gamma_{K_L} \frac{m}{E_K} t_o) = 0.01 $.
However due to the fact that  $\Gamma_{K_L} \ll  \Gamma_{K_S}$
the expected limits for $\delta_{K_L}$ and $\delta_{K_S}$
don't differ strongly, namely
\begin{equation}
\begin{split}
&\delta_{K_L}   \leq 1.3 \cdot 10^{-17}~GeV \\
&\delta_{K_S}   \leq  0.33 \cdot 10^{-17}~GeV
\end{split}
\end{equation}

\par In the case of a signal observation,  several methods could be used to cross-check the result. For instance, 
to test whether the  signal is due to the HCAL non-hermeticity or not, one could perform  measurements with different HCAL thicknesses, i.e. with  one, two, three,  and four consecutive HCAL modules.  In this case the expected background level could be obtained by extrapolating the results to an infinite HCAL thickness.  
The evaluation of the signal and background  could also be  obtained from the results of measurements at different 
 beam energies. 
 \par The signal from $K_{S,L} - K^M_{S,L}$ oscillations can be cross-checked by 
 \begin{itemize}
 \item modifying the length of the $K^0$ decay volume
 \item changing the air pressure in the decay volume 
 \item changing the energy of the beam
 \end{itemize}

\section{Conclusion}
Due to their specific properties, neutral kaons  are  one of the most interesting probes of physics beyond the standard model  from both 
 theoretical and  experimental viewpoints. 
 The  decays $\ksl$ have never been experimentally tested. In the Standard Model their  branching ratios for the decay into two neutrinos are predicted to be extremely small, $Br(K_{S,L} \to \nu \bar{\nu}) \lesssim 10^{-16}$.  Thus, observation of  $\ksl$ decays would unambiguously signal the presence of new physics. 
\par  We consider   the $K_L \to invisible$ decay in several natural extensions of the SM, such as the 2HDM, 2HDM and light neutral scalar field $\phi $, and dark mirror matter model.  Using constraints from the experimental value for  the  $Br(K^+ \rightarrow \pi^+ \nu \bar{\nu})$ we find that the  $\ksl$ decay  branching ratio could be in the   region $Br(\ksl)\simeq 10^{-8}- 10^{-6}$, 
which is  experimentally accessible allowing to test new-physics scales well above 100 TeV. In some scenarios
 the  bound $Br(K_{S,L} \to \nu \bar{\nu}) \lesssim 10^{-16}$ can be avoided, as  in the model with the massive  right-handed neutrino and scalar 
$\phi$-particle. All this makes  these decay a  powerful clean  probe of  new physics, that is complementary to  other  rare $K$ decay channels. Additionally, in the case of observation,  $\ksl$ decays could  influence  the Bell-Steinberger analysis of the $K^0-\overline{K}^0$ system. 
\par The results obtained provide a strong motivation for a sensitive search for these decay modes in a near future experiment similar to the one proposed in \cite{Gninenko}. 
 We  briefly discussed such a search  with the  pion and kaon beams available  at  the CERN PS and SPS, which would be also capable of improving sensitivity  
 for invisible decays of  $\eta, \eta'$ and  other neutral mesons.
If such decays  exist, they could be observed by looking for events with a 
striking signature: the total disappearance of the beam energy in a fully  hermetic hadronic  calorimeter.
 A feasibility study of the experimental setup shows that this unique signature  allows for searches of  $\kslinv$ decays with a sensitivity in the branching ratio $Br(\ksl) \lesssim 10^{-7} -10^{-5}$, and 
$\eta, \eta' \to invisible$ decays with a sensitivity a few  orders of magnitude beyond the present experimental limits.

These results could be obtained with a detector that is  optimized for  
several of properties,  namely, i) the intensity and purity of the primary pion and kaon beams,  ii)  
high-efficiency  tagging of the $K^0$ production, and  iii) a high level 
of hermeticity in the hadronic  calorimeter  system are of importance.
Large amounts of high-energy hadrons and high background suppression are crucial to improving the sensitivity of the search. To obtain the best limits,  a compromise should be found between the background level and the energy and intensity  of the beam.  
\par Finally, we note that the presented analysis gives an illustrative order of magnitude for the sensitivity of the proposed experiment and may be strengthened  by more detailed 
simulations of the  experimental setup.

\section*{Acknowledgments}

We would like to  thank  our colleagues from the NA64 Collaboration for their interest to this work. In particular, we are grateful to  L.~Molina Bueno,  A.~Celentano,  P.~Crivelli,  S.~Donskov, M.~Kirsanov,  S.~Kuleshov, V.~Lyubovitskij,  D. Peshekhonov, V.~Poliakov, V.~Samoylenko,  A.~Toropin, and A.~Zevlakov  for useful discussions. 
\vskip0.3cm

{\it Note added.} - Recently, we became aware of a related work \cite{bes25}.

\end{document}